MGR INŻ. MARCIN KAWALEROWICZ

# CLASSIFICATION OF AUTOMATIC SOFTWARE BUILD METHODS

Summary: The process of creating working software from source code and other components (like libraries, database files, etc.) is called "software build". Apart from linking and compiling, it can include other steps like automated testing, static code analysis, documentation generation, deployment and other. All that steps can be automated using a build description of some sort (e.g. script). This article classifies the automatic software build processes beginning at build script and reaching the various types of continuous integration.

Chapter 1. Classification of automatic software build methods

Software build process at a basic level consists of the translating the source code into working software. The translated code is often unit tested. More mature software build processes include more steps like complex testing, code metrics, code analysis, deployment [1] and other. So the build process contains a set of repetitive manual tasks that integrate the various components (source code, libraries, data files, etc.) into software. The tasks need to be executed in certain chronological order. Output of every task can vary from execution to execution, depending on the state the source code is in. So the next task to execute can vary depending on the output of the previous task. As a series of repetitive tasks the software build process done by humans is error prone [2]. To eliminate the uncertainty of the "human factor" the build process is often recommended to be automated [3].

Figure 1 shows the build automation diagram containing the classification of the automatic software build processes. First of all the build processes can be done on-demand or in a continual way. The on-demand build automation execution was named a "Levered" execution because there is a need to use some hypothetical lever to start the process. This lever could be a command line command execution or a mouse click in development environment. As such the levered execution can be viewed as a commanded integration. Someone or something needs to execute a command to integrate the components into working software. The build process can be automated using a build script: a textual (seldom binary) description of the steps needed to be executed to build a software.

On the other side of the first differentiation we have a continual integration of the software components. It often uses the same build mechanism (script) as in the levered method but it is set into motion by other events. The continual build automation is commonly known as continuous integration [4] (or simply by its abbreviation CI).

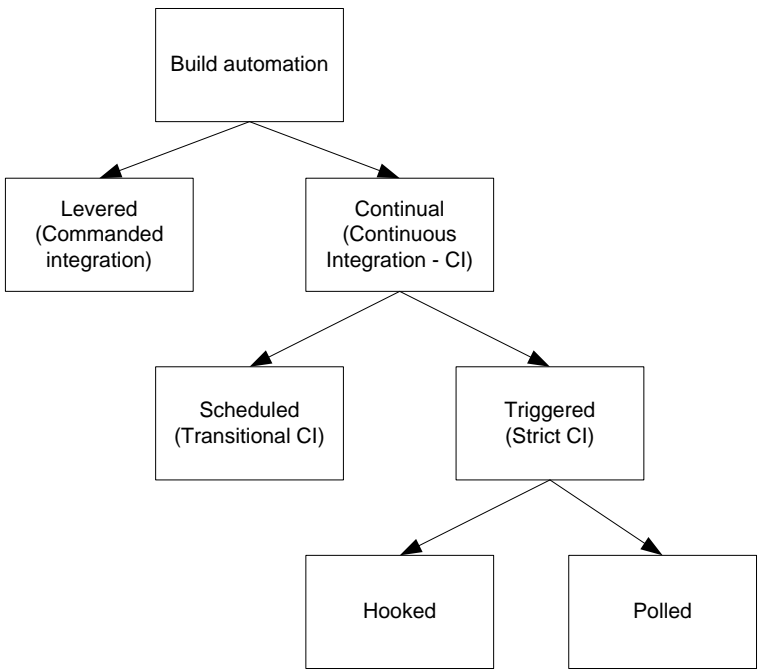

Figure 1 Classification of automatic software build methods

The continuous integration build automation method is used in development teams that use a central code base repository [5]. Such repository contains the "most current" version of source code and other components. It is the place where the every developers passes the code he writes. Continual builds are commonly executed on dedicated computers known as continuous integration servers.

Depending on the event that starts the integration in the continual way, we can single out the scheduled and triggered builds [6]. Scheduled builds are executed according to a schedule. Most common type of scheduled build is so called

"night build" – a build executed in the time of least human software development activity (which is usually at night). This type of build automation was classified as a "transitional continuous integration", because it can be viewed as a way to accomplish more mature "strict continuous integration". In the strict type of build automation the build process needs to be triggered by an external event. The trigger setting the build into motion is most often the change in a source code base. Any change in the code repository leads to full software integration process. In a opposition to scheduled build automation this type is called triggered build automation.

In a strict continuous integration environment the changes in the code repository can be detected either by polling the changes form the code base or by hooking the repository to the continuous integration process. In the first case the CI server calls periodically to the central repository to check if the source code changed since last integration. If so the changes are pulled from repository and the integration is performed. This method of triggering the build is called "polled". On the other hand in the "hooked" method the repository itself informs the CI server about the change in the source code. A notification procedure of some kind is hooked to the repository that triggers the integration on the CI server.

The strict continuous integration method is sometimes lessened if the continuous integration server is not able to perform the integration after every change in the central source repository. Under a heavy load (time intensive build or/and large number of developers passing the code to the repository) the build queue grows and the feedback from the process is reaching the developers comes late – which is not acceptable in continuous integration [1]. In such scenario there is a "quiet time period" introduced on the CI server. After every build the server waits given time and gathers the repository changes executing the build not after every change but after a period of time. This type of behaviour cannot be considered as strict CI. While still triggered it should be included to the transitional CI methods because the build is not executed after every change in the central repository.

Conclusions

This paper described the build automation classification dividing the automated build on the levered and continual builds. It introduces continuous integration division to transitional (where build is not executed after every change; e.g. scheduled) and strict (where build is triggered after every change in the central repository). This paper described also methods of triggering the builds using polling and hooking.

December 2011 / Januar 2012